\documentclass[rmp,twocolumn,showpacs,superscriptaddress]
{revtex4-2}
\pdfoutput=1

\usepackage[normalem]{ulem}
\usepackage{graphicx}
\usepackage{amsmath}
\usepackage{amssymb}
\usepackage{latexsym}
\usepackage{fancyhdr}
\usepackage{array}
\usepackage{amsfonts}
\usepackage{mathrsfs}
\usepackage{mathtools}
\usepackage{color}
\usepackage{verbatim}
\usepackage{physics}
\usepackage[caption=false]{subfig}
\usepackage{natbib}
\usepackage{enumitem}
\usepackage{bbold}
\usepackage[colorlinks=true,urlcolor=blue,citecolor=blue,linkcolor=blue]{hyperref}

\usepackage{float}
\usepackage{xcolor}
\usepackage{color,soul}

\begin{document}

\title{Relative facts do not exist. Relational Quantum Mechanics is Incompatible with Quantum Mechanics. Response to the critique by Aur\'elien Drezet.}

\author{Jay Lawrence}
\affiliation{Department of Physics and Astronomy,
Dartmouth College, Hanover, NH 03755, USA}

\author{Marcin Markiewicz}
\affiliation{International Centre for Theory of Quantum Technologies (ICTQT),
University of Gdansk, 80-308 Gdansk, Poland}

\author{Marek \.Zukowski}
\affiliation{International Centre for Theory of Quantum Technologies (ICTQT),
University of Gdansk, 80-308 Gdansk, Poland}

\date{\today}

\begin{abstract}
In this comment we answer to the recent critique of our article [arXiv:2208.11793] about Relational Quantum Mechanics (RQM) by Aur\'elien Drezet [arXiv:2209.01237]. Here we point out that our critical analysis of RQM was precisely based on the most recent formulation of RQM, and that the theses found in the critique are based on neither RQM assumptions nor on our arguments.

\end{abstract}

\maketitle

%\section{Introduction}

We recently posted an article on the arXiv entitled 
``Relative facts do not exist:  Relational Quantum 
Mechanics is Incompatible with Quantum Mechanics'' \cite{Lawrence22}.
We prove this assertion by deriving a Greenberger Horne 
Zeilinger (GHZ) contradiction within a Wigner-Friend 
scenario.  The contradiction arises for the characteristic 
(GHZ) correlations among measurement outcomes on 
three-qubit systems (to which we shall refer below). We 
emphasise that our critique pertains to the most up-to-date 
presentation of Relational Quantum Mechanics (RQM) published 
in 2021 by its founder Carlo Rovelli \cite{Rovelli.21}. This
presentation has recently been  incorporated as chapter 43 in  
``The Oxford Handbook of the History of Quantum Interpretations'' 
published on 7 June 2022 by Oxford University Press.

A critique of our article \cite{Lawrence22} was posted shortly 
thereafter by Aur\'elien Drezet, entitled ``In defense of 
Relational Quantum Mechanics: A note on (above title)'' 
\cite{Drezet22}.  Far from defending  RQM, however, 
Drezet misrepresents it, as well as our paper.  Based on 
his misunderstandings, the critique of our paper is 
unfounded.  Here we discuss several specific instances, in 
order of appearance in his paper:
\begin{enumerate}
\item A repeated misgiving regarding RQM is first revealed
in paragraph 2 of Drezet's paper, where he ``reminds us'' 
that:\\
``\textit{... in RQM the main issue concerns the 
interpretation of the full wavefunction $\ket{\Psi_{SO}}$
involving observer (O) and observed system (S).  In RQM 
the fundamental object \underline{relative to (O)} is not
$\ket{\Psi_{SO}}$ but the reduced density matrix}
\begin{equation}
   \hat{\rho}_{S|O}^{red.} = \hbox{Tr}_O[\hat{\rho}_{SO}] =
   \hbox{Tr}_O[|\Psi_{SO} \rangle \langle \Psi_{SO}|]\textrm{''}.
\label{rho1}
\end{equation}
This statement is wrong for at least two reasons:  First, the reduced density
matrix is {\it not} the fundamental object relative to ($O$).  Quoting Rovelli 
\cite{Rovelli.21}, [RQM has] ``\textit{an ontology based on facts (or events), 
not quantum states}.'' Second, Drezet’s formula (1) provides
an incorrect description of what the observer sees in a measurement.
Not only in RQM, but also in ordinary quantum mechanics, the observer
sees a single unambiguous outcome (in RQM, this is called a ``fact’’).  The 
reduced density matrix, in contrast, represents an ensemble average.  This 
is not what the observer sees in a single measurement.

\item Based on the above  (inappropriate) use of the reduced density matrix, 
Drezet compares our GHZ correlation equations with his
own, which of course show less correlation and do not pose a 
contradiction.  The two cases are Eqs. (14) and (15) for the 
Friend (portrayed here by Alice), and later, Eqs. (19)
and (27) for Wigner (played here by Bob). To be more explicit, he states [immediately prior to Eq. (14)] ``\textit{Due to decoherence 
i.e., entanglement with the environment (Alice) we have 
lost coherence and correlations between spins.}'' This flatly contradicts the
description of Wigner-Friend scenarios provided by Rovelli, dating back to the original 1996  work \cite{Rovelli.96}, and more comprehensively in his
recent article \cite{Rovelli.21}, 
in which relative facts are realized by Alice and, by 
definition, have not yet been converted into stable facts 
by decoherence. To illustrate this point, let us again cite Rovelli's description of the Wigner-Friend scenario (\cite{Rovelli.21}, page 3):\\

\begin{itemize}\item
\textit{For instance, in the Wigner’s friend scenario, the friend
interacts with a system and a fact is realised with respect
to the friend. But this fact is not realised with respect
to Wigner, who was not involved in the interaction, and
the probability for facts with respect to Wigner (realised
in interactions with Wigner) still includes interference effects.}\end{itemize}
\item Drezet concludes his commentary on Alice's RQM measurements with 
the sentence (following Eq. 15): ``\textit{The actualization of 
measurements in RQM is a debatable issue and we will not consider 
this problem here.}'' This statement further misrepresents RQM:  
first, the term “actualization” is not defined or used in RQM.  
Instead, referring to the above quote of Rovelli, ``\textit{a 
fact is *realized* with respect to the friend}'' (here Alice).  
This is conventional usage in RQM, and it’s meaning is not 
debatable.  The subject of measurements in RQM is discussed more 
broadly by Di Biagio and Rovelli \cite{Biagio.22}, and, as far as  
we know, this presentation is not generally considered to be
debatable.  

It may be worth noting that we reviewed the ``rules of RQM” in 
Sec. II of our paper, before implementing them in Sec. III to set 
up the Wigner-Friend scenario.  We did not make up our own rules.

\item In criticizing the treatment of Bob's RQM measurements on the 
compound system $S \otimes A$, Drezet again introduces the reduced 
density matrix (as mentioned in item 1) to (erroneously) describe 
the situation relative to Bob.  Using this, he finds Eq. (19) on 
p. 5, which shows no (GHZ) correlations in three of the four cases 
where we find them.  What is more, he misrepresents our
argumentation by saying, that "\textit{they consider that Bob only 
measures one of the 3 qubits belonging to $SA$}". In fact, in our 
scenario, Bob performs unitary interactions, which in RQM lead to relative facts, on each of the three subsystems 
$S_i \otimes A_i$ (see top paragraph, right column of p. 4 in \cite{Lawrence22}). 
The ordering of these RQM measurements is immaterial, since the 
entangling processes commute, so that Bob's RQM measurements may be 
regarded as simultaneous.  We then consider the state (13) [or (20) 
in \cite{Drezet22}], {\it not} to describe further RQM measurements,
but only to find deterministic relations between the relative facts
established in previous measurements, as exhibited by eigenvalues of the involved observables. 

%More specifically, in (13) and its permutations, we apply Bob's \marek{
%unitary transformations, 
%each acting on only one compound system $S_i \otimes A_i$, to
%generate three different GHZ ``Hilbert space vectors'' which are 
%all isomorphic to the states $\ket{GHZ}_{SA}$ and $\ket{GHZ}_{SAB}$ 
%that were generated by RQM measurements.} The relative facts must
%exhibit all of the GHZ correlations thus obtained.

\item Drezet challenges our assertion that relative facts are 
non-contextual. He argues that we introduce non-contextuality
arbitrarily,
misrepresenting us with the statement (p. 6, ten lines from the
bottom), ``\textit{since the 3 operations … are acting ‘locally’ 
on only one of the subsystems $SA_m$ (at a time) their meaning 
should be non-contextual and absolute.}’’ We do not say this, nor do we assume 
it.\footnote{On the contrary, local measurements
can be used to demonstrate contextuality.}
In brief, our paper demonstrates that
Alice's relative facts, which (according to RQM)  come into 
existence with Alice's RQM measurements, are non-contextual 
hidden variables  with respect to the context of Bob's future
measurements.  In contradistinction with other hidden
variables theories, they do not exist in RQM at the stage of 
the preparation of the experiment, or equivalently, of the 
quantum state of the system ($S$) in question. Recall that, 
as mentioned in our paper, we regard a hidden variable as 
a notion, variable, value, or whatever which is not an 
element of quantum formalism. Relative facts are definitely 
not a part of quantum formalism.
\end{enumerate}

As a final remark let us emphasize once more, that in our work we do not compare values of relative facts which are realised with respect to different observers. This no comparison rule, or "relativity of comparisons" axiom, is one of the main features postulated in \cite{Biagio.22}\footnote{However, please note that this axiom is withdrawn, and replaced by "cross-perspective links" axiom, which does not rule out comparisons of values... \cite{Rovelli2203}. This effectively puts relative facts as counterfactuals, exactly of the form discussed in "step two" of our letter \cite{Zukowski.21} }. Instead we utilise constraints on products of them, which are demanded by ordinary quantum mechanics. If one rejects the assumption, that relative facts should follow the same constraints as corresponding eigenvalues of quantum observables, one immediately looses any relation with quantum mechanics.
%, without need for further more involved argumentation. }

\section*{Acknowledgements}
This work is partially supported by  Foundation for Polish 
Science (FNP), IRAP project ICTQT, contract no. 2018/MAB/5, 
co-financed by EU Smart Growth Operational Programme. MZ 
thanks Michail Skotiniotis for a discussion on the link of 
all that with non-contextual hidden variables. MZ also 
thanks the late Jarek Pykacz for years of discussions 
on the subject of hidden variables.

\bibliography{Rovelli}

\end{document}